\documentclass[12pt]{article}
\usepackage{amssymb,amsmath,epsfig}

\begin{document}

\title{\bf Cosmic Acceleration and Brans-Dicke Theory}

\author{M. Sharif \thanks{msharif.math@pu.edu.pk} and Saira Waheed
\thanks{smathematics@hotmail.com}\\
Department of Mathematics, University of the Punjab,\\
Quaid-e-Azam Campus, Lahore-54590, Pakistan.}

\date{}

\maketitle
\begin{abstract}
This paper is devoted to study the accelerated expansion of the
universe by exploring the Brans-Dicke parameter in different eras.
For this purpose, we take FRW universe model with viscous fluid
(without potential) and Bianchi type I universe model with
barotropic fluid (with and without potential). We evaluate
deceleration parameter as well as Brans-Dicke parameter to explore
cosmic acceleration. It is concluded that accelerated expansion of
the universe can also be achieved for higher values of the
Brans-Dicke parameter in some cases.
\end{abstract}
{\bf PACs numbers: 04.50.Kd, 98.80.-k};

\section{Introduction}

The accelerated expansion of the observable universe is one of the
most conspicuous and recent achievement in modern cosmology. This
expansion with positive cosmic acceleration has been confirmed by
many astronomical experiments such as Supernova (Ia) $^{1,2)}$,
WMAP $^{3)}$, SDSS $^{4)}$, galactic cluster emission of X-rays
$^{5)}$, large scale structure $^{6)}$, weak lensing $^{7)}$ etc.
These results lead to the conclusion that our universe is
spatially flat.

The positive cosmic acceleration of the universe has been
motivated by a mysterious exotic matter having large negative
pressure, known as dark energy. Although, General Relativity (GR)
is an excellent theory to explain the gravitational effects but it
is unable to describe the present cosmic acceleration and the
reality of dark energy. In order to explain the nature of this
mysterious finding, various models including Chaplygin gas,
phantom, quintessence, cosmological constant etc. have been
constructed $^{8,9)}$. However, none of these models is very
successful.

The exploration of scalar-tensor theories of gravity as modified
theories of gravity has received much attention due to their vast
implications in cosmology $^{10-14)}$. Brans-Dicke (BD) theory of
gravity, a special case of scalar-tensor theories, is one of the
most viable theories for this purpose. It is the general
deformation of GR satisfying weak equivalence principle, in which
gravity effects are mediated by the metric tensor and the scalar
field $^{15)}$. This provides a direct coupling of the scalar
field to geometry. Brans-Dicke theory is compatible with both
Mach's principle $^{16)}$ and Dirac's large number hypothesis
$^{17)}$. One of the salient features of this theory is that the
gravitational coupling constant, being the inverse of spacetime
scalar field, varies with time. In order to fulfill the solar
system experiment constraints, the value of the generic
dimensionless BD parameter $\omega$ should be very large, i.e.,
$\omega\geq40,000$ $^{18,19)}$.

Brans-Dicke theory is a successful theory which can tackle many
outstanding cosmological problems like inflation, quintessence,
late time behavior of the universe, coincidence problem, cosmic
acceleration $^{11)}$ etc. There are different versions of BD
theory available in literature $^{20,21)}$. Singh and Rai $^{22)}$
investigated various BD cosmological models and showed that the
Bianchi models are very effective in explaining the evolution of
the universe for perfect fluid. Bermann $^{23)}$ discussed
different models of the universe with constant deceleration
parameter based on the variation law of Hubble parameter.
Bertolami and Martins $^{11)}$ found that the accelerated
expansion of the universe could be obtained with large $|\omega|$
and potential $\phi^{2}$ without considering the positive energy
condition. Sen and Sen $^{24)}$ showed that the dissipative
pressure could support the late time accelerated expansion of the
universe. Banerjee and Pavon $^{12)}$ found that the present
accelerated expansion could be obtained without restoring a
cosmological constant or quintessence matter for FRW model.

Sahoo and Singh $^{25)}$ explored the observed accelerated expansion
of the present universe in this theory for FRW model. The same
authors $^{13)}$ found exact solutions in different eras of the
universe and discussed the possibilities for obtaining cosmic
acceleration, inflation and deceleration for these solutions. Sen
and Seshadri $^{26)}$ investigated the role of positive power law
potential on the accelerated expansion of the universe. They
concluded that self-interacting potential can derive the accelerated
expansion in the perfect fluid background with small negative values
of BD parameter. Reddy and Rao $^{27)}$ found axially symmetric
perfect fluid cosmological model in this theory. In order to
investigate the present accelerated expansion of the universe and
different stages of the cosmic evolution, a lot of work has been
done using Bianchi models in GR and scalar tensor theories
$^{28-32)}$. In a recent paper, Chakraborty and Debnath $^{14)}$
have investigated cosmic acceleration in this theory for FRW model.
They have shown that the accelerated expansion of the universe with
higher values of $\omega$ can be achieved only for closed model.

In this paper, we explore the role of BD parameter on the cosmic
acceleration by using spatially flat models in the presence of
different fluids. The paper is organized as follows. In the next
section, we formulate the field equations of generalized BD theory
with self-interacting potential. Section \textbf{3} provides the
field equations for FRW model in the presence of a viscous fluid.
Here we discuss models for both constant as well as varying bulk
viscosity coefficient. In section \textbf{4}, we formulate the field
equations in the presence of the barotropic fluid for the Bianchi
type I (BI) universe model. This section explores all possible
choices of the BD parameter $\omega$ and the self-interacting
potential $V(\phi)$. Section \textbf{5} investigate the
observational limit of gravitational constant for the constructed
models. Finally, we discuss the results in the last section.

\section{Brans-Dicke Field Equations}

Brans and Dicke $^{15)}$ proposed a scalar-tensor theory known as
Brans-Dicke theory of gravity that was based on the pioneering work
of Jordan. A modified version of this theory is the generalized BD
theory in which the BD parameter no longer remains a constant
rather, it turns out to be a function of the scalar field. The
action for generalized BD theory with self-interacting potential in
Jordan frame $^{20,21)}$ is given by
\begin{equation}\label{1}
S=\int d^{4}x\sqrt{-g}[\phi
R-\frac{\omega(\phi)}{\phi}\phi^{,\alpha}\phi_{,\alpha}-V(\phi)+L_{m}];~~\alpha=0,1,2,3,
\end{equation}
where BD parameter $\omega(\phi)$ is the modified form of the
original BD parameter $\omega,~V(\phi)$ denotes the
self-interacting potential and $L_{m}$ represents the matter part
of the Lagrangian. Here we have taken $8\pi G_{0}=c=1$. Taking
variation of this action with respect to the metric tensor
$g_{\mu\nu}$ and the scalar field, we obtain the following BD
field equations $^{14)}$
\begin{eqnarray}\nonumber
G_{\mu\nu}&=&\frac{\omega(\phi)}{\phi^{2}}[\phi_{,\mu}\phi_{,\nu}
-\frac{1}{2}g_{\mu\nu}\phi_{,\alpha}\phi^{,\alpha}]+\frac{1}{\phi}[\phi_{,\mu;\nu}
-g_{\mu\nu}\Box\phi]\\\label{2}&-&\frac{V(\phi)}{2\phi}g_{\mu\nu}
+\frac{T_{\mu\nu}}{\phi},\\\nonumber
\Box\phi&=&\frac{T}{3+2\omega(\phi)}-\frac{1}{3+
2\omega(\phi)}[2V(\phi)-\phi\frac{dV(\phi)}{d\phi}]\\\label{3}
&-&\frac{\frac{d\omega(\phi)}{d\phi}}{3+2\omega(\phi)}\phi_{,\mu}\phi^{,\mu},
\end{eqnarray}
where $T=g^{\mu\nu}T_{\mu\nu}$ denotes trace of the
energy-momentum tensor and $\Box$ represents the d'Alembertian
operator. Equation (\ref{3}) is called wave equation for the
scalar field. Notice that BD theory is reducible to GR if $\omega
\rightarrow \infty$ and the scalar field becomes a constant
$^{33)}$. However, it is not true in general. In papers $^{30,
34)}$, it has been pointed out that BD theory does not always go
over to GR in the limit $\omega \rightarrow \infty$ for the case
of exact solutions. In this limit, GR could be recovered only if
the trace of the energy-momentum tensor $T^{(m)}$ describing all
fields other than BD scalar field does not vanish, i.e.,
$T^{(m)}\neq 0$ $^{34-37)}$. For $T^{(m)}=0$, the BD solutions do
not correspond to respective GR solutions. The Palatini metric
$f(R)$ gravity and the metric $f(R)$ gravity are obtained by
substituting $\omega=-3/2$ and $\omega=0$ respectively $^{38)}$.

\section{Cosmic Acceleration and FRW Model}

In this section, we investigate cosmic acceleration by exploring
the BD parameter. For this purpose, we consider FRW model with
viscous fluid. In particular, we discuss two cases according to
bulk viscosity is constant or variable. The line element for FRW
model is given by
\begin{equation}\label{4}
ds^{2}=dt^{2}-a^{2}(t)[\frac{dr^{2}}{1-kr^{2}}+r^{2}(d\theta^{2}+\sin^{2}\theta
d\phi^{2})],
\end{equation}
where $a(t)$ is the scale factor and $k=-1,0,+1$ indicate open,
flat and closed universe model respectively. We assume that the
universe is filled with a viscous fluid given by
\begin{equation}\label{5}
T_{\mu\nu}=(\rho+P_{eff})u_{\mu}u_{\nu}-P_{eff}g_{\mu\nu},
\end{equation}
where $\rho$ is the energy density, $u^{\mu}$ is the four-vector
velocity satisfying the relation $u_{\mu}u^{\nu}=1$ and $P_{eff}$
represents the effective pressure defined by
\begin{equation*}
P_{eff}=P_{I}+P_{vis}.
\end{equation*}
Here $P_{I}$ denotes the isotropic pressure and $P_{vis}$
represents the pressure due to viscosity. The bulk viscous
pressure is defined by Eckart's expression in terms of fluid
expansion scalar and is given by $P_{vis}=-\xi u^{\mu}_{;\mu}$
$^{39)}$, where $\xi=\xi(t,\rho)$ represents the bulk viscosity
coefficient. For FRW model, the viscous pressure is found to be
$P_{vis}=-\frac{3\xi \dot{a}}{a}$ and hence the effective pressure
becomes
\begin{equation}\label{5a}
P_{eff}=P_{I}-3\xi H,
\end{equation}
where $H=\frac{\dot{a}}{a}$ denotes the Hubble parameter. The
corresponding field equations (\ref{2}) turn out to be
\begin{eqnarray}\label{6}
&&\frac{\dot{a}^{2}+k}{a^{2}}+\frac{\dot{a}\dot{\phi}}{a\phi}
-\frac{\omega}{6}\frac{\dot{\phi}^{2}}{\phi^{2}}=\frac{\rho}{3\phi},\\\label{7}
&&\frac{2\ddot{a}}{a}+\frac{\dot{a}^{2}+k}{a^{2}}+\frac{\omega}{2}
\frac{\dot{\phi}^{2}}{\phi^{2}}+\frac{2\dot{a}\dot{\phi}}{a\phi}+\frac{\ddot{\phi}}{\phi}=
\frac{-P_{eff}}{\phi},
\end{eqnarray}
where dot denotes the derivative with respect to time. The
corresponding wave equation becomes
\begin{equation}\label{8}
\ddot{\phi}+3H\dot{\phi}=\frac{\rho-3P_{eff}}{2\omega
+3}-\frac{\dot{\omega}\dot{\phi}}{2\omega +3}.
\end{equation}
Here we have taken $V(\phi)=0$.

Equation of state provides a relation between isotropic pressure
and energy density and is given by
\begin{equation}\label{9}
P=\gamma\rho,
\end{equation}
where $\gamma$ is the equation of state (EoS) parameter. The
values of $\gamma=-1,~0,~1/3,~1$ represent vacuum dominated, dust,
radiation dominated era and massless scalar field respectively.
The continuity equation for the viscous fluid (\ref{5}) can be
written as
\begin{equation}\label{10}
\dot{\rho}+3H(\rho+P_{eff})=0.
\end{equation}
One can assume the standard expression for bulk viscosity, i.e.,
$\xi= \xi_{0}\rho^{n}$, where $n$ is a non-negative constant and
$\xi_{0}>0$. Different possible values of $n$ are available in
literature $^{40-43)}$, out of which two choices $n=1$ and $n=3/2$
correspond to the radiative and string dominated fluids
respectively. However, more realistic models can be obtained for
$0\leq n\leq 1/2$. Here we would like to evaluate $\rho$ by
solving the continuity equation (\ref{10}) for the following two
cases:
\begin{itemize}
\item Constant bulk viscosity, i.e., $\xi=\xi_{0}$ (for $n=0$).
\item Variable bulk viscosity, i.e., $\xi=\xi(t,\rho)$ with
$n=1/2,1$.
\end{itemize}
In both cases, we choose $k=0$, i.e, flat FRW model.

\subsection{Constant Bulk Viscosity Coefficient}

The energy conservation equation (\ref{10}), in terms of constant
bulk viscosity, can be written as
\begin{equation}\label{a}
\dot{\rho}+3H(1+\gamma)\rho=9\xi_{0}H^{2},
\end{equation}
where we have used the EoS given by Eq.(\ref{9}). We assume that
the scale factor $a(t)$ has the form of expanding solution (power
law form)
\begin{equation}\label{11}
a(t)=a_{0}t^\alpha,\quad\alpha>0,
\end{equation}
where $a_0$ is the present value of the scale factor. The
deceleration parameter is given by
\begin{equation}\label{11a}
q=-(1+\frac{\dot{H}}{H^{2}}).
\end{equation}
Notice that the deceleration parameter $q$ suggests $\alpha>1$ for
cosmic acceleration. Equation (\ref{a}) leads to
\begin{equation}\label{12}
\rho(t)=\frac{9\xi_{0}\alpha^{2}}{[-1+3\alpha(1+\gamma)]}t^{-1}
+\rho_{0}a_{0}^{-3(1+\gamma)}t^{-3\alpha(1+\gamma)}.
\end{equation}
The scalar field can be found from Eq.(\ref{8}) by taking
$\omega(t)=\omega_{0}$ (constant) as follows
\begin{equation*}
\phi(t)=\frac{(1-3\gamma)\rho_{0}a_{0}^{-3(1+\gamma)}t^{[2-3\alpha(1+\gamma)]}}
{(3+2\omega_{0})(1-3\alpha\gamma)[2-3\alpha(1+\gamma)]}
+\frac{3\xi_{0}(1-4\alpha)t}{(3+2\omega_{0})[1-3\alpha(1+\gamma)]}.
\end{equation*}
This equation suggests that the scalar field can be taken in the
power law form when the scale factor is given in the expanding
form.

We would like to discuss the time dependent BD parameter $\omega$
which satisfies the field equations as well as wave equation. For
this purpose, we assume a simple form of power law for the scalar
field
\begin{equation}\label{11b}
\phi(t)=\phi_{0}t^\beta,
\end{equation}
where $\phi_{0}$ is the present value of the scalar field and
$\beta$ is any non-zero constant. The field equation (\ref{6}) can
be re-arranged in the following form
\begin{equation}\label{12a}
\frac{\dot{a}}{a}=-\frac{\dot{\phi}}{2\phi}\pm\sqrt{\frac{\Omega(t)
\dot{\phi}^{2}}{12\phi^{2}}+\frac{\rho}{3\phi}},
\end{equation}
where $\Omega(t)=2\omega(\phi(t))+3$. Using Eqs.(\ref{11}) and
(\ref{11b}) in Eq.(\ref{12a}), it follows that
\begin{equation}\label{13}
\rho(t)=3\phi_{0}t^{\beta}[\frac{(2\alpha
+\beta)^{2}}{4t^{2}}-\frac{\Omega(t)\beta^{2}}{12t^{2}}].
\end{equation}
The comparison of Eqs.(\ref{12}) and (\ref{13}) yields
\begin{equation}\label{14}
\Omega(t)=\frac{3}{\beta^{2}}(2\alpha+\beta)^{2}-\frac{36\xi_{0}\alpha^{2}}
{\phi_{0}\beta^{2}[3\alpha(1+\gamma)-1]}t^{(1-\beta)}-\frac{4\rho_{0}a_{0}^{-3(1+\gamma)}
t^{[-3\alpha(1+\gamma)-\beta+2]}}{\phi_{0}\beta^{2}}.
\end{equation}
The corresponding expression for $\omega(t)$ will become
\begin{equation}\label{15}
\omega(\phi(t))\approx
\omega(t)=\frac{-18\xi_{0}\alpha^{2}t^{(1-\beta)}}{\phi_{0}\beta^{2}[3\alpha(1+\gamma)-1]}
-\frac{2\rho_{0}a_{0}^{-3(1+\gamma)}t^{[2-\beta-3\alpha(1+\gamma)]}}{\phi_{0}\beta^{2}}.
\end{equation}
Here we have considered the time dependent terms only.

In order to check the consistency of these solutions with the wave
equation, we substitute these values in (\ref{8}). This leads to
the following two consistency relations
\begin{eqnarray}\label{d1}
&&\beta[4(\beta-1+3\alpha)+\beta(1-3\gamma)+2[2-3\alpha(1+\gamma)-\beta]]=0,\\\nonumber
&&4\alpha(\beta-1+3\alpha)+\alpha(1-3\gamma)\beta+[3\alpha(1+\gamma)
-1]\beta+2\alpha(1-\beta)=0.\\\label{e}
\end{eqnarray}
Equation (\ref{d1}) implies that either $\beta=0$ or
$\beta=-2\alpha$ while Eq.(\ref{e}) is satisfied for either
$\beta=-2\alpha$ or $\alpha=1/6$. For cosmic acceleration
$\alpha=1/6$ is not an interesting value and so we ignore it. When
$\beta=0$, the BD parameter yields
$\omega(t)\longrightarrow-\infty$ and the scalar field becomes a
constant, i.e., $\phi=\phi_{0}$. This leads to GR, so it is not
the interesting case. For $\beta=-2\alpha,~\omega$ takes the form
\begin{equation}\label{16}
\omega(t)=-\frac{9\xi_{0}t^{(1+2\alpha)}}{2\phi_{0}[3\alpha(1+\gamma)-1]}
-\frac{\rho_{0}a_{0}^{-3(1+\gamma)}t^{[2-\alpha(1+3\gamma)]}}{2\phi_{0}\alpha^{2}}.
\end{equation}
The power law expression for the scalar field turns out to be
\begin{equation*}
\phi(t)=\phi_{0}t^{-2\alpha}.
\end{equation*}
In the following, we evaluate the BD parameter at different epochs
of the universe.

For \textbf{vacuum dominated era} ($\gamma=-1$), the BD parameter
is
\begin{equation}\label{17}
\omega(t)=\frac{9\xi_{0}}{2\phi_{0}}t^{(1+2\alpha)}-\frac{\rho_{0}}{2\phi_{0}
\alpha^{2}}t^{2(1+\alpha)},
\end{equation}
while in the \textbf{radiation dominated era} ($\gamma=1/3$), it
becomes
\begin{equation}\label{18}
\omega(t)=\frac{9\xi_{0}}{2\phi_{0}(1-4\alpha)}t^{(1+2\alpha)}
-\frac{\rho_{0}a_{0}^{-4}t^{2(1-\alpha)}}{2\phi_{0}
\alpha^{2}}.
\end{equation}
The BD parameter in the \textbf{matter dominated era} or the dust
case ($\gamma=0$) takes the form
\begin{equation}\label{19}
\omega(t)=\frac{9\xi_{0}}{2\phi_{0}(1-3\alpha)}t^{(1+2\alpha)}
-\frac{\rho_{0}a_{0}^{-3}t^{(2-\alpha)}}{2\phi_{0}
\alpha^{2}}.
\end{equation}
In the \textbf{massless scalar field era} ($\gamma=1$), this turns
out to be
\begin{equation}\label{20}
\omega(t)=\frac{9\xi_{0}}{2\phi_{0}(1-6\alpha)}t^{(1+2\alpha)}
-\frac{\rho_{0}a_{0}^{-6}t^{(2-4\alpha)}}{2\phi_{0}
\alpha^{2}}.
\end{equation}
Finally, the BD parameter for the \textbf{present time}, $t=t_{0}$,
can be calculated from dust case, i.e., matter with negligible
pressure. Equation (\ref{19}) leads to the present value of the BD
parameter $\omega_{0}$ given by
\begin{equation}\label{21}
\omega_{0}=-\frac{9\xi_{0}}{2(3\alpha-1)}-\frac{1}{2\alpha^{2}}.
\end{equation}
Here we have normalized the constants, i.e.,
$\phi_{0}=a_{0}=t_{0}=1,~\rho_{0}=0$ and $\alpha\geq 1$ which is
consistent with Eq.(\ref{12a}). The minimum value of $\omega_{0}$
is
\begin{equation*}
\omega_{0}=-\frac{9\xi_{0}}{4}-\frac{1}{2}.
\end{equation*}

Clearly the minimum value of $\omega_{0}$ depends on the value of
constant bulk viscosity coefficient $\xi_{0}$. In the BD theory,
the gravitational coupling constant and the scalar field density
should be positive in the present universe which can be achieved
for $\omega>-3/2$ $^{12)}$. In our case, the bulk viscosity
coefficient must have $\xi_{0}<4/9$ with $\alpha>1$ for the
consistency purpose. The present observational range for the
deceleration parameter is $-1<q_{0}<0$ $^{1,2)}$ which restricts
$\alpha>1$. The more general form of the model for the present
universe can be obtained by taking $\alpha=1+\epsilon,~\epsilon>0$
(for small values of $\epsilon$) given by
\begin{eqnarray}\nonumber
\omega(t)=-\frac{9\xi_{0}t^{(3+2\epsilon)}}{2(2+3\epsilon)}
-\frac{t^{(1-\epsilon)}}{2(1+\epsilon)^{2}},\quad
\phi=\phi_{0}t^{-2(1+\epsilon)},\quad a(t)=a_{0}t^{(1+\epsilon)}.
\end{eqnarray}
\begin{figure}
\centering \epsfig{file=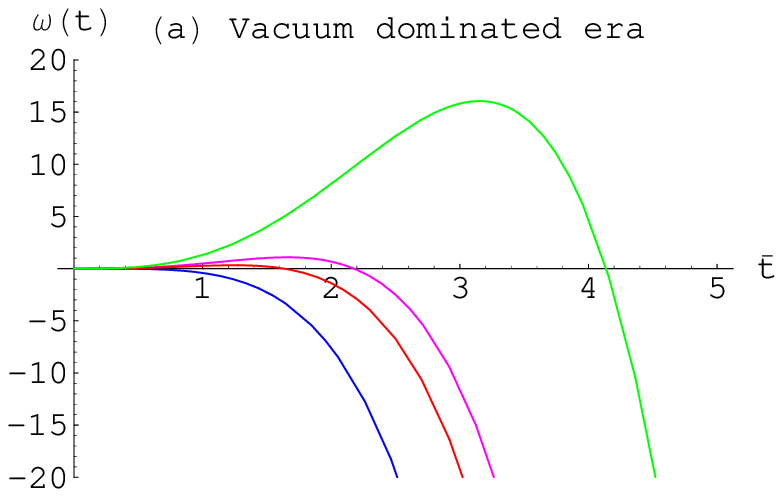,width=.45\linewidth}
\epsfig{file=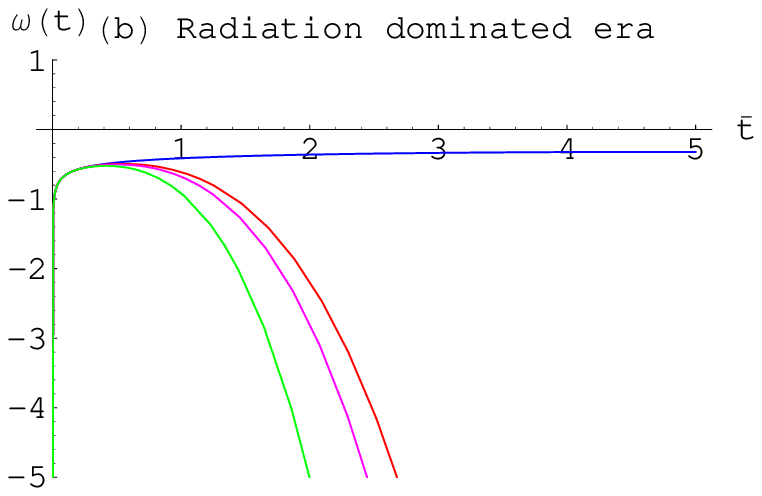,width=.45\linewidth} \caption{Plots show the
graph of $\omega(t)$ versus cosmic time $t$ for (a) $\gamma=-1$
and (b) $\gamma=1/3$ with $\alpha=1.1$ and varying values of
$\xi_{0}$ as follows: blue, $\xi_{0}=0.0001$; red, $\xi_{0}=0.15$;
pink, $\xi_{0}=0.2$; green, $\xi_{0}=0.38$.}
\end{figure}

Now we discuss the BD parameter for vacuum as well as matter
dominated eras. In the vacuum dominated era, the graphs indicate
that $\omega(t)$ is a decreasing function starting from zero for
$0<\xi_{0}\leq 0.11$. For $\xi_{0}>0.11$, the graphs correspond to
increasing function but after some particular points, they again
become decreasing function as shown in Figure \textbf{1}. Thus, for
this range of constant viscosity $\xi_{0}$ with $\alpha>1$, it is
possible to achieve the cosmic acceleration with positive values of
$\omega(t)$. In all other eras of the universe, $\omega(t)$ is a
decreasing function of time with smaller negative values. For
$0<\alpha<1$, in the radiation dominated era, $\omega(t)$ is a
decreasing function and the universe undergoes to decelerated
expansion. Thus, $\xi_{0}$ plays the role to control the time
dependence of $\omega(t)$. In the radiation, matter dominated eras
and massless scalar field, the BD parameter approaches to $-\infty$
for $\alpha=1/6,~1/3$ and $1/4$. For the cosmic acceleration, we
must have $\alpha>1$, hence these values are not interesting.

\subsection{Variable Bulk Viscosity Coefficient}

For the sake of simplicity, we take $n=1/2$, i.e.,
$\xi(t,\rho)=\xi_{0}\rho^{1/2}(t)$. Using this value of bulk
viscosity coefficient along with Eq.(\ref{11}) in (\ref{10}), it
follows that
\begin{equation*}
\dot{\rho(t)}+[\frac{3\alpha}{t}(1+\gamma)-\frac{9\xi_{0}\alpha^{2}}{t^{2}}]\rho(t)^{1/2}=0.
\end{equation*}
This yields the following solution
\begin{equation}\label{100}
\rho(t)=\left[\frac{9\xi_{0}\alpha^2}{t[3\alpha(1+\gamma)-2]}+\rho_{0}t^{-3\alpha(1+\gamma)/2}\right]^2,
\end{equation}
where $\rho_{0}$ is an integration constant. Comparing this
equation with Eq.(\ref{13}), we obtain
\begin{equation*}
\Omega(t)=3\left(\frac{2\alpha+\beta}{\beta}\right)^2-\frac{4t^2}{\phi_{0}\beta^{2}t^{\beta}}
\left[\frac{9\xi_{0}\alpha^2}{t[3\alpha(1+\gamma)-2]}+\rho_{0}t^{-3\alpha(1+\gamma)/2}\right]^2.
\end{equation*}
The corresponding BD parameter will become
\begin{equation}\label{c}
\omega(t)=-\frac{2t^2}{\phi_{0}\beta^{2}t^{\beta}}
\left[\frac{9\xi_{0}\alpha^2}{t[3\alpha(1+\gamma)-2]}+\rho_{0}t^{-3\alpha(1+\gamma)/2}\right]^2.
\end{equation}
For consistency of this solution with the wave equation, we
substitute all these values in the wave equation (\ref{8}) which
leads to
\begin{equation*}
\alpha=1/3,\quad\beta=-2\alpha,\quad\gamma=1.
\end{equation*}
For $\beta=-2\alpha$, we obtain
\begin{equation}\label{d}
\omega(t)=-\frac{t^{2(1+\alpha)}}{2\phi_{0}\alpha^2}
\left[\frac{9\xi_{0}\alpha^2}{t(3\alpha(1+\gamma)-2)}+\rho_{0}t^{-3\alpha(1+\gamma)/2}\right]^2.
\end{equation}
The choice $\alpha=1/3$ is not feasible for obtaining cosmic
acceleration while $\gamma=1$ corresponds to massless scalar field
which is discussed below. Now we evaluate BD parameter for
different eras.

In the \textbf{vacuum dominated era}, the BD parameter is
\begin{equation}\label{102}
\omega(t)=-\frac{t^{2(1+\alpha)}}{2\phi_{0}\alpha^2}
\left[\frac{-9\xi_{0}\alpha^2}{2t}+\rho_{0}\right]^2
\end{equation}
while for the \textbf{radiation dominated era}, it turns out to be
\begin{equation}\label{103}
\omega(t)=-\frac{t^{2(1+\alpha)}}{2\phi_{0}\alpha^2}
\left[\frac{9\xi_{0}\alpha^2}{t(4\alpha-2)}+\rho_{0}t^{-2\alpha}\right]^2.
\end{equation}
The BD parameter in the \textbf{matter dominated era} is
\begin{equation}\label{104}
\omega(t)=-\frac{t^{2(1+\alpha)}}{2\phi_{0}\alpha^2}
\left[\frac{9\xi_{0}\alpha^2}{t(3\alpha
-2)}+\rho_{0}t^{-3\alpha/2}\right]^2.
\end{equation}
\begin{figure}
\centering \epsfig{file=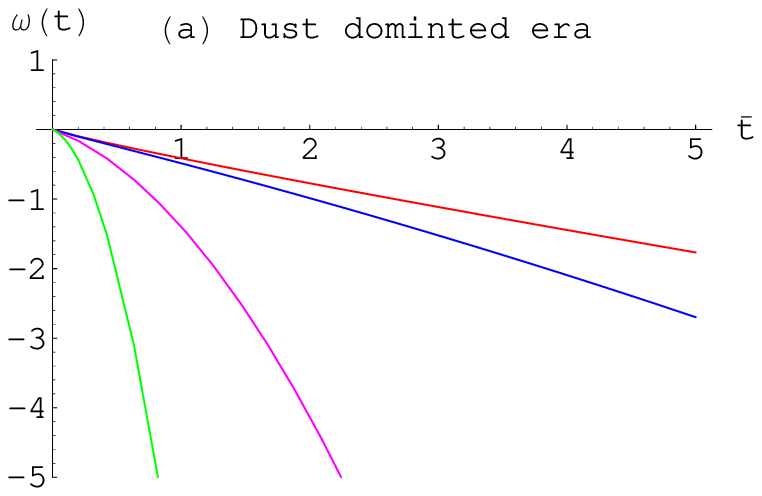,width=.45\linewidth}
\epsfig{file=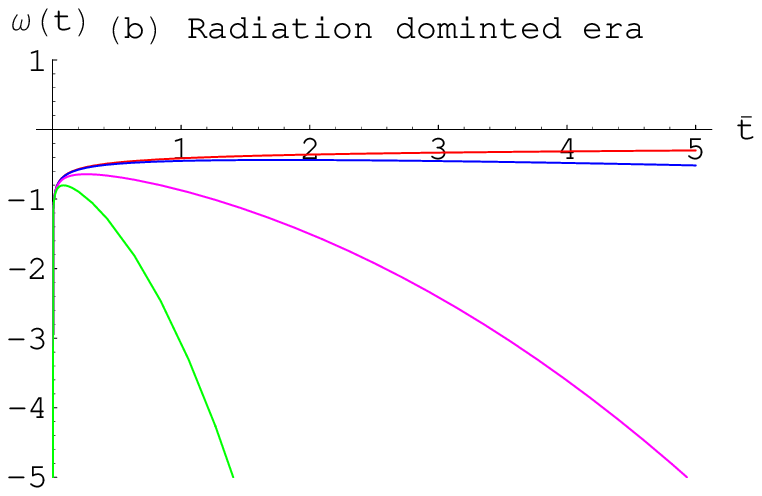,width=.45\linewidth} \caption{Plots show the
graphs for $\omega(t)$ versus cosmic time $t$ for $\alpha=1.1$,
(a) $\gamma=0$ and (b) $\gamma=1/3$ with varying values of
$\xi_{0}$ as follows: red, $\xi_{0}=0.0001$; green,
$\xi_{0}=0.38$; blue, $\xi_{0}=0.01$; pink, $\xi_{0}=0.1$.}
\end{figure}
\begin{figure}
\centering \epsfig{file=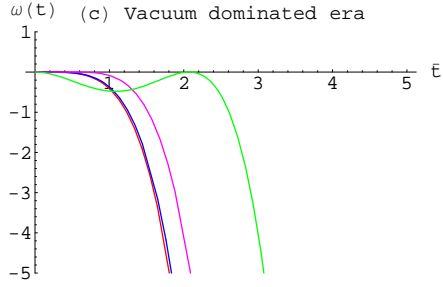,width=.45\linewidth}
\caption{Plots shows the graphs for $\omega(t)$ versus cosmic time
$t$ for $\alpha=1.1$ and $\gamma=-1$ with varying values of
$\xi_{0}$ as follows: red, $\xi_{0}=0.0001$; green,
$\xi_{0}=0.38$; blue, $\xi_{0}=0.01$; pink, $\xi_{0}=0.1$.}
\end{figure}
For the \textbf{massless scalar field era}, this is given by
\begin{equation}\label{105}
\omega(t)=-\frac{t^{2(1+\alpha)}}{2\phi_{0}\alpha^2}
\left[\frac{9\xi_{0}\alpha^2}{t(6\alpha-2)}+\rho_{0}t^{-3\alpha}\right]^2.
\end{equation}
The expressions for $\omega(t)$ correspond to decreasing function
as $-t^{2}$ for increasing values of viscosity coefficient
$\xi_{0}$ and $-1\leq\gamma\leq 1$ except for vacuum dominated
era. This gives rise to accelerated expansion of the universe for
$\alpha>1$ as shown in Figures \textbf{2} and \textbf{3}. For
$\alpha=1/2,~3/2$ and $\alpha=1/3$, the BD parameter approaches to
$-\infty$. Here in the matter dominated era, $\alpha=3/2$ lies in
the range $\alpha>1$ allowed for the accelerated expansion of the
universe.

Now we discuss the radiative fluid case ($n=1$). Here we take
$\xi(t,\rho)=\xi_{0}\rho(t)$. Consequently, the continuity
equation yields
\begin{equation}\label{105}
\rho(t)=\rho_{0}t^{-3\alpha(1+\gamma)}\exp{(\frac{-9\xi_{0}\alpha^{2}}{t})}.
\end{equation}
The BD parameter $\omega(t)$ turns out to be
\begin{equation*}
\omega(t)=-\frac{2\rho_{0}
\exp{(\frac{-9\xi_{0}\alpha^{2}}{t})}}{\phi_{0}\beta^{2}}t^{[(2-3\alpha(1+\gamma)-\beta]}.
\end{equation*}
Here $\beta=0$ and $\beta=-2\alpha$ are the corresponding
consistency relations. The choice of $\beta=0$ provides no
interesting insights while $\beta=-2\alpha$ leads to the following
expression
\begin{equation}\label{101}
\omega(t)=-\frac{\rho_{0}
\exp{(\frac{-9\xi_{0}\alpha^{2}}{t})}}{2\phi_{0}\alpha^{2}}t^{[2-\alpha(1+3\gamma)]}.
\end{equation}
For the radiation dominated era, the BD parameter takes the form
\begin{equation}\label{103}
\omega(t)=-\frac{\rho_{0}
\exp{(\frac{-9\xi_{0}\alpha^{2}}{t})}}{2\phi_{0}\alpha^{2}}t^{2(1-\alpha)}.
\end{equation}
We see that the coefficient of viscosity appears only in the
exponential function. Thus, in the radiation dominated era, for
small values of $\xi_{0}$ and
$\alpha>1,~\exp(-9\xi_{0}\alpha/t)\longrightarrow 1$, providing
small negative values of $\omega(t)$ as shown in Figure
\textbf{4}. If $\xi_{0}\longrightarrow \infty$ with $\alpha>1$,
then $\exp(-9\xi_{0}\alpha/t)\longrightarrow 0$ which means
$\omega(t)\longrightarrow 0$. Thus this model may correspond to
that of the metric $f(R)$ gravity. However, it is not physically
possible.  Also, in this case, for $0<\alpha<1$, the values of
$\omega(t)$ are constrained within the range $-3/2<\omega(\phi)<0$
which shows that the universe undergoes to the decelerated phase.
\begin{figure}
\centering \epsfig{file=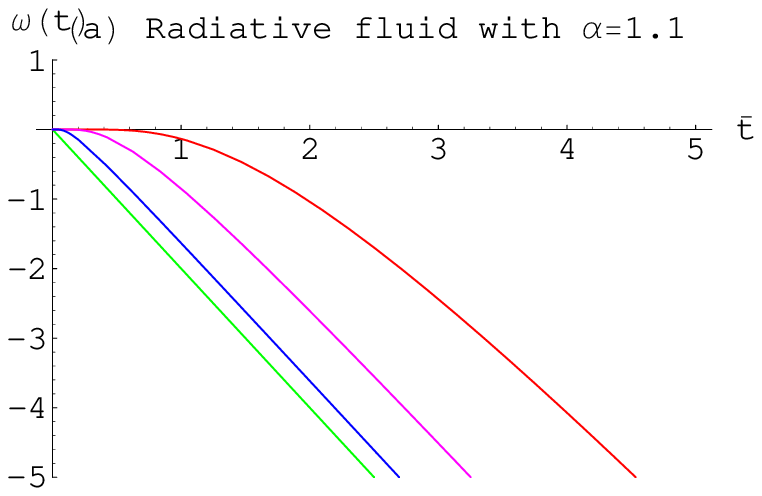,width=.45\linewidth}
\epsfig{file=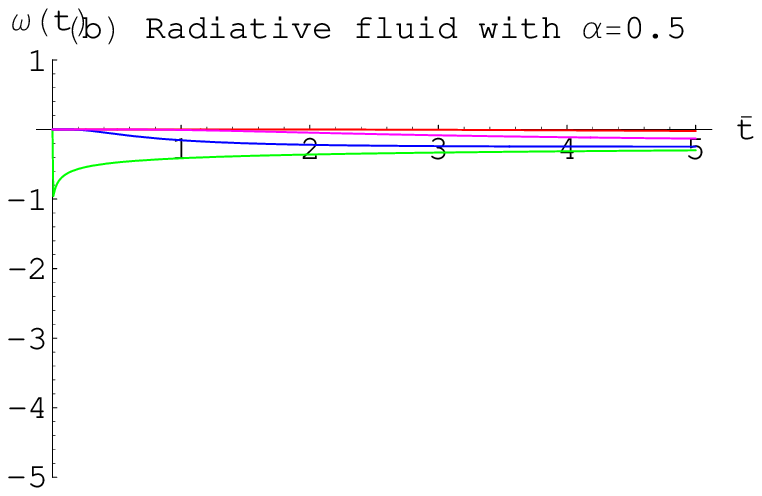,width=.45\linewidth} \caption{Plots show the
graphs for $\omega(t)$ versus cosmic time $t$ for (a) $\alpha=1.1$
and (b) $\alpha=0.5$ with varying values of $\xi_{0}$ as follows:
green, $\xi_{0}=0.0001$; red, $\xi_{0}=1.2$; blue, $\xi_{0}=0.09$;
pink, $\xi_{0}=0.38$.}
\end{figure}

\section{Cosmic Acceleration with Barotropic fluid and Bianchi I Universe Model}

Here we investigate expansion of the universe by using LRS Bianchi
type \textit{I} model in the barotropic fluid background. The line
element of Bianchi type I universe model is described by $^{44)}$
\begin{equation}\label{51}
ds^{2}=dt^{2}-A^{2}(t)dx^{2}-B^{2}(t)(dy^{2}+dz^{2}),
\end{equation}
where $A$ and $B$ are the scale factors. This model has one
transverse direction $x$ and two equivalent longitudinal
directions $y$ and $z$. Assume that matter contents of the
universe are described by the perfect fluid given by
\begin{equation}\label{x}
T_{\mu\nu}=(\rho+P)u_{\mu}u_{\nu}-Pg_{\mu\nu}.
\end{equation}
The corresponding field equations (\ref{2}) and (\ref{3}) can be
written as
\begin{eqnarray}\label{53}
&&\frac{2\dot{A}\dot{B}}{AB}+\frac{\dot{B}^{2}}{B^{2}}=\frac{\rho}{\phi}+\frac{\omega(\phi)}{2}
\frac{\dot{\phi}^{2}}{\phi^{2}}+\frac{V(\phi)}{2\phi}-(\frac{\dot{A}}{A}
+2\frac{\dot{B}}{B})\frac{\dot{\phi}}{\phi},\\\label{54}&&
2\frac{\ddot{B}}{B}+\frac{\dot{B}^{2}}{B^{2}}=-\frac{P}{\phi}-\frac{\omega(\phi)}{2}
\frac{\dot{\phi}^{2}}{\phi^{2}}-2\frac{\dot{B}}{B}\frac{\dot{\phi}}{\phi}-\frac{\ddot{\phi}}{\phi}
+\frac{V(\phi)}{2\phi},\\\label{55}&&
\frac{\ddot{B}}{B}+\frac{\ddot{A}}{A}+\frac{\dot{A}\dot{B}}{AB}=-\frac{P}{\phi}-\frac{\omega(\phi)}{2}
\frac{\dot{\phi}^{2}}{\phi^{2}}-\frac{\ddot{\phi}}{\phi}+\frac{V(\phi)}{2\phi}-(\frac{\dot{A}}{A}
+\frac{\dot{B}}{B})\frac{\dot{\phi}}{\phi}.
\end{eqnarray}
The wave equation is
\begin{equation}\label{56}
\ddot{\phi}+(\frac{\dot{A}}{A}+2\frac{\dot{B}}{B})\dot{\phi}=\frac{\rho-3P}{2\omega(\phi)
+3}-\frac{[2V(\phi)-\phi\frac{dV(\phi)}{d\phi}]}{2\omega(\phi)
+3}-\frac{\frac{d\omega(\phi)}{d\phi}\dot{\phi}^{2}}{2\omega(\phi)+3}.
\end{equation}
For this model, the average scale factor and the mean Hubble
parameter are
\begin{equation*}
a^3(t)=AB^{2},\quad
H(t)=\frac{1}{3}(\frac{\dot{A}}{A}+2\frac{\dot{B}}{B}).
\end{equation*}
The energy conservation equation for energy-momentum tensor given
in Eq.(\ref{x}) will be
\begin{equation}\label{57}
\dot{\rho}+(\frac{\dot{A}}{A}+2\frac{\dot{B}}{B})(\rho+P)=0.
\end{equation}

We assume that the universe is filled with barotropic fluid. The
barotropic EoS $^{14)}$ is given by
\begin{equation*}
P=\gamma\rho;\quad(-1\leq\gamma\leq1).
\end{equation*}
The expansion scalar for Bianchi type I model is given by
\begin{equation*}
\theta=u^{a}_{;a}=\frac{\dot{A}}{A}+2\frac{\dot{B}}{B}
\end{equation*}
while the shear scalar is
\begin{equation*}
\sigma=\frac{1}{\sqrt{3}}(\frac{\dot{A}}{A}-\frac{\dot{B}}{B}).
\end{equation*}
It is given $^{45)}$ that for spatially homogeneous metric, the
normal congruence to homogeneous expansion yields the ratio
$\frac{\sigma}{\theta}$ as constant i.e., "expansion scalar
$\theta$ is proportional to shear scalar $\sigma$". This physical
condition leads to the following relation between the scale
factors
\begin{equation}\label{58}
A=B^{m},
\end{equation}
where $m\neq1$ is any positive constant (for $m=1$, it reduces to
flat FRW model). In literature $^{44-49)}$, this condition has
been widely used to find exact cosmological models. Using this
assumption in Eq.(\ref{57}), it follows that
\begin{equation*}
\dot{\rho}+(1+\gamma)(m+2)\frac{\dot{B}}{B}\rho(t)=0
\end{equation*}
which yields
\begin{equation}\label{59}
\rho(t)=\rho_{0}B^{-(1+\gamma)(m+2)}.
\end{equation}
Now we discuss the various possible choices for $\omega(\phi)$ and
$V(\phi)$.

\subsection{Model without potential $V(\phi)=0$}

We take the following two cases according to $\omega$ is constant
and $\omega=\omega(\phi)$.

\subsubsection{Case (i)}

First we take BD parameter as a constant, i.e.,
$\omega(\phi)=\omega_{0}$. For the solution of the field
equations, we consider the power law as follows
\begin{equation}\label{60}
B(t)=b_{0}t^{\alpha},\quad\alpha\geq 0.
\end{equation}
Using Eqs.(\ref{58}), (\ref{60}) and the mean Hubble parameter $H$,
the deceleration parameter can be written as
\begin{equation*}
q=-[1-\frac{3}{\alpha(m+2)}].
\end{equation*}
Notice that $q<0,~q=0$ and $q>0$ indicate an accelerated expansion,
uniform expansion and the decelerating phase of the universe
respectively. Thus, for accelerated expansion of the universe, we
must have the following condition on $\alpha$
\begin{equation}\label{65}
\alpha>\frac{3}{(m+2)};\quad m\neq 1.
\end{equation}
Substituting Eqs.(\ref{58}) and (\ref{60}) in (\ref{56}), the
scalar filed becomes
\begin{equation}\label{61}
\phi(t)=\frac{(1-3\gamma)\rho_{0}b_{0}^{-(m+2)(1+\gamma)}t^{[2-\alpha(1+\gamma)(m+2)]}}
{(3+2\omega_{0})[1-\alpha\gamma(m+2)][2-\alpha(1+\gamma)(m+2)]}.
\end{equation}
The BD parameter is obtained from the field equations
(\ref{53})-(\ref{55}) as
\begin{eqnarray}\nonumber
\omega_{0}&=&-\frac{1}{(1-\gamma)}[\frac{(m+3)\alpha(\alpha-1)}
{[2-\alpha(m+2)(1+\gamma)]^2}+\alpha^{2}\frac{[(m^{2}+1)
+2\gamma(2m+1)]}{[2-\alpha(m+2)(1+\gamma)]^2}\\\label{62}
&+&\frac{\alpha[m+3+2\gamma(m+2)]}
{[2-\alpha(m+2)(1+\gamma)]}+\frac{[1-\alpha(m+2)(1+\gamma)]}{[2-\alpha(m+2)(1+\gamma)]}].
\end{eqnarray}
\begin{figure}
\centering \epsfig{file=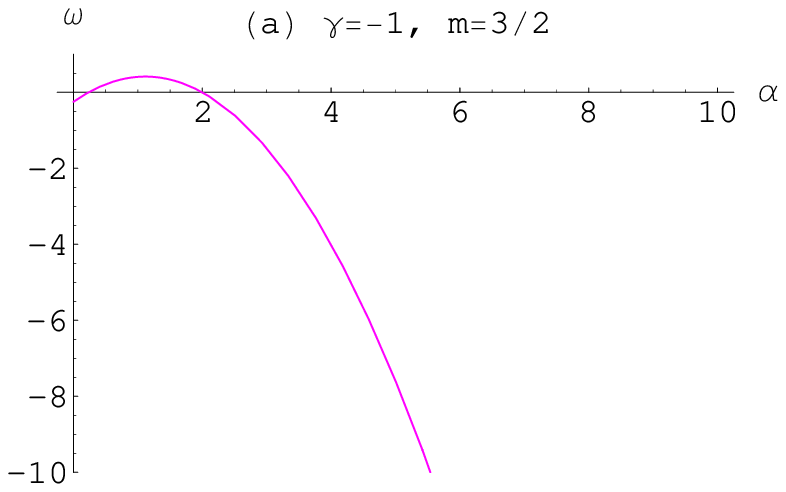,width=.45\linewidth}
\epsfig{file=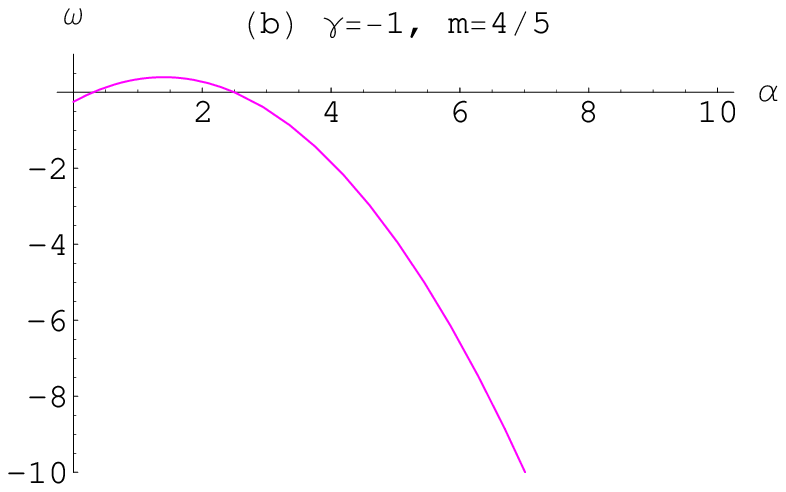,width=.45\linewidth} \caption{Plot of
$\omega_{0}$ versus $\alpha$ for (a) $m=3/2$, (b) $m=4/5$ and
$\gamma=-1$. The corresponding ranges for $\alpha$ are
$\alpha>6/7$ and $\alpha>15/14$.}
\end{figure}
\begin{figure}
\centering \epsfig{file=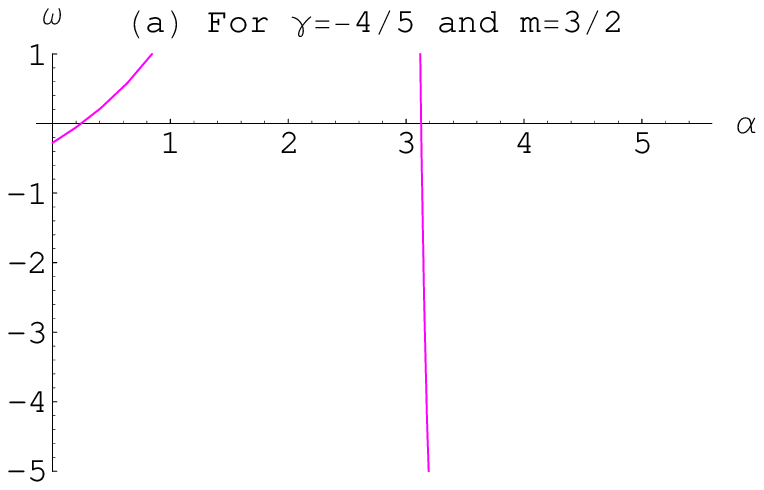,width=.45\linewidth}
\epsfig{file=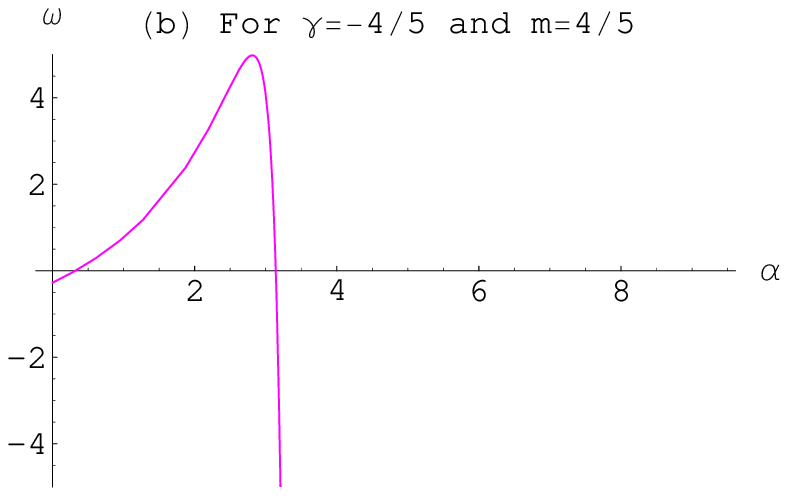,width=.45\linewidth} \caption{Plot of
$\omega_{0}$ versus $\alpha$ for (a) $m=3/2$, (b) $m=4/5$ and
$\gamma=-4/5$. The corresponding ranges for $\alpha$ are
$\alpha>6/7$ and $\alpha>15/14$.}
\end{figure}
\begin{figure}
\centering \epsfig{file=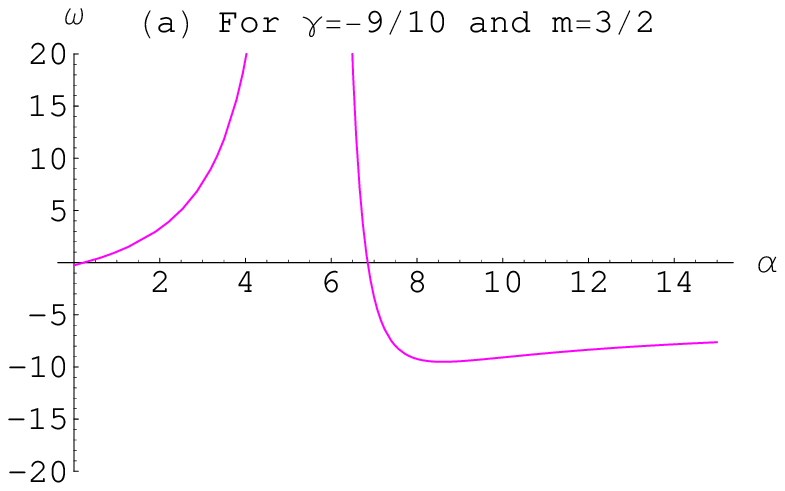,width=.45\linewidth}
\epsfig{file=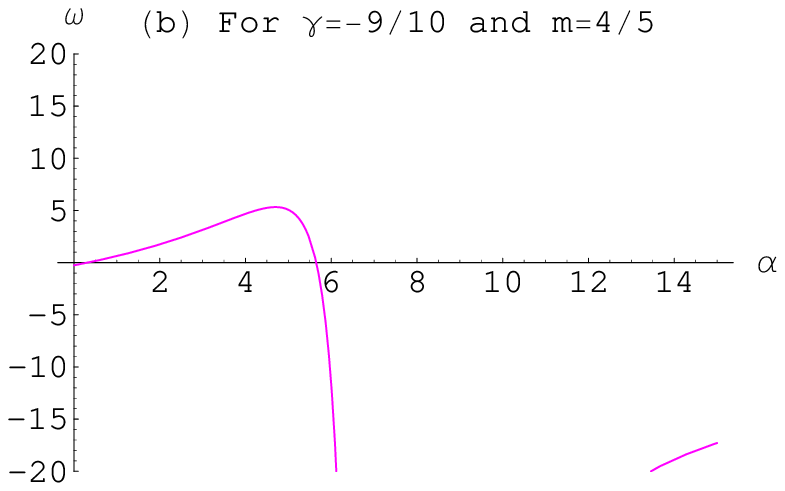,width=.45\linewidth} \caption{Plot of
$\omega_{0}$ versus $\alpha$ for (a) $m=3/2$, (b) $m=4/5$ and
$\gamma=-9/10$. The corresponding ranges for $\alpha$ are
$\alpha>6/7$ and $\alpha>15/14$.}
\end{figure}

For massless scalar field $\gamma=1$, we have
$\omega\longrightarrow -\infty$, which leads to GR. We have seen
that the BD parameter depends upon the parameters $\alpha,~\gamma$
and $m$. These parameters are constrained using some physical
conditions. The possible ranges for $m$ are $0<m<1$ and $m>1$ and
$\gamma$ is allowed for $-1\leq\gamma\leq1$. The deceleration
parameter constraints $\alpha$ such that $\alpha>3/(m+2)$. By
taking different possible choices for these parameters, it can be
seen that BD parameter takes small negative values as well as
positive values for $-1\leq\gamma<0$ as shown in Figures
\textbf{5-7}. This gives rise to cosmic acceleration for this
range of $\gamma$. We would like to mention here that for ceratin
ranges of $\alpha$ allowed for cosmic acceleration and
$-1\leq\gamma<0,~\omega_{0}$ can take larger values which would be
compatible with the solar system experiment constraints.

Solving Eq.(\ref{62}) for $\alpha$, we obtain the following
quadratic equation
\begin{eqnarray}\nonumber
&&\alpha^{2}[(m+2)^2(1+\gamma)^2[(\gamma-1)\omega_{0}-2]-(m+3)-[m^{2}+1+2\gamma(2m+1)]
\\\nonumber&&+(m+2)(1+\gamma)[m+3+2\gamma(m+2)]]
+\alpha[(m+2)(1+\gamma)[-4(\gamma-1)\omega_{0}\\\label{d}
&&+6](m+3)-2[m+3+2\gamma(m+2)]]+4[\omega_{0}(\gamma-1)-1]=0
\end{eqnarray}
which provides two roots. These values for $m=1/2$ and $\gamma=0$
(present universe) are given by
\begin{eqnarray}\label{63}
\alpha=\frac{23/2+10\omega_{0}\pm\sqrt{-15/4-6\omega_{0}}}{17+25/2\omega_{0}}.
\end{eqnarray}
Since $-2\leq\omega_{0}\leq-3/2$ is the observed range for cosmic
acceleration, so the choice of $\omega_{0}=-5/3$ leads to
following values of $\alpha$
\begin{equation*}
\alpha_{1}=16/23,\quad\alpha_{2}=2.
\end{equation*}
Here $\alpha_1$ gives $q>0$, hence we leave it while $\alpha_2=2$
yields $q<0$ leading to accelerating expansion. Also, it yields
$\phi(t)=t^{-3}$ which provides positive coupling constant. Since
in our case, $\phi(t)$ is decreasing more rapidly as compared to
$\phi(t)=t^{-2}$ $^{11)}$ and $\phi(t)=t^{-5/2}$ $^{12)}$,
therefore it corresponds to greater rate of accelerated expansion
of the universe.

\subsubsection{Case (ii)}

In this case, the BD parameter is not constant rather it is a
function of $\phi$. Using Eqs.(\ref{11}), (\ref{58}),
(\ref{53})-(\ref{55}) and (\ref{60}), the BD parameter can be
written as
\begin{eqnarray}\nonumber
\omega(\phi)&=&\frac{1}{\beta^{2}}[\frac{(3m-m^{2}-2)}{2}\alpha^{2}
+\frac{(m+3)\alpha}{2}+\frac{(m+1)\alpha\beta}{2}-\beta^{2}+\beta]
\\\label{67}&-&\frac{1}{\beta^{2}}[\rho_{0}b_{0}^{-(m+2)(1+\gamma)}
(1+\gamma)\phi^{\frac{-\alpha(m+2)(1+\gamma)
-\beta+2}{\beta}}\phi_{0}^{\frac{-\alpha(m+2)(1+\gamma)-2}{\beta}}].
\end{eqnarray}
Substituting this value in Eq.(\ref{56}), we obtain the following
consistency relation
\begin{equation}\label{68}
\beta=-\frac{(m+2)\alpha(1+\gamma)}{2};\quad m\neq1
\end{equation}
This shows that $\beta$ remains negative for all $0<m<1,~m>1$,
$\alpha>3/(m+2)$ and $-1\leq\gamma\leq1$. The consistency of this
solution with the dynamical equations, i.e., each term in the
dynamical equations should have the same time dependence, results
in another constraint given by
\begin{equation}\nonumber
\beta=2-\alpha(m+2)(1+\gamma).
\end{equation}
Using this value of $\beta$ in Eq.(\ref{68}), it can be seen that
the parameter $\beta$ is restricted to $-2$. Now we discuss the BD
parameter and cosmic acceleration in different phases of the
universe by using this value of $\beta$. The expressions for BD
parameter in matter and radiation dominated eras with $\beta=-2,
~\alpha>6/5$ and $m=1/2$ turn out to be
\begin{eqnarray}\nonumber
\omega(\phi)=\frac{1}{4}[-\frac{3\alpha^{2}}{8}+\frac{\alpha}{4}-6]
-\frac{1}{4}\phi^{(-2+\frac{5\alpha}{4})},\\\nonumber
\omega(\phi)=\frac{1}{4}[-\frac{3\alpha^{2}}{8}+\frac{\alpha}{4}-6]
-\frac{1}{3}\phi^{(-2+\frac{5\alpha}{3})}.
\end{eqnarray}
\begin{figure}
\centering \epsfig{file=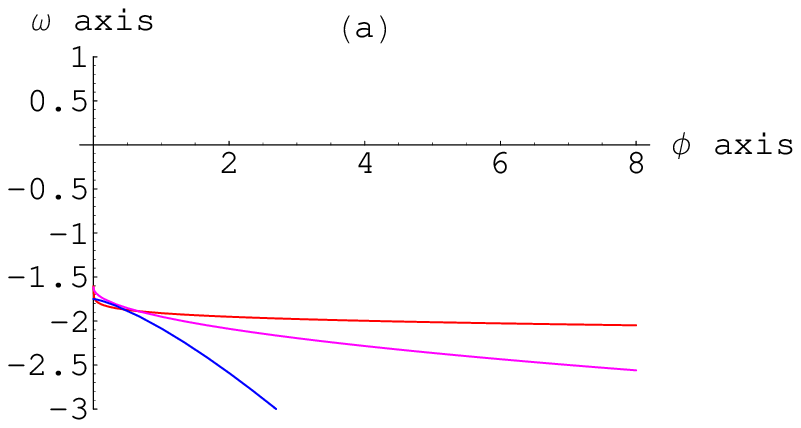,width=.45\linewidth}
\epsfig{file=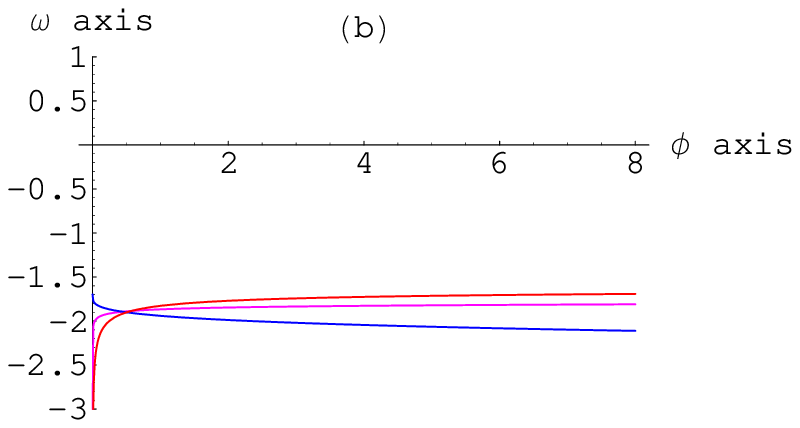,width=.45\linewidth} \caption{Plot of
$\omega(\phi)$ versus $\phi$ for $m=1/2$ and (a) $\gamma=1/3$ and
(b) $\gamma=0$ with $\alpha>6/5$ as follows: red, $\alpha=1.3$;
blue, $\alpha=2$; pink, $\alpha=1.5$.}
\end{figure}
\begin{figure}
\centering \epsfig{file=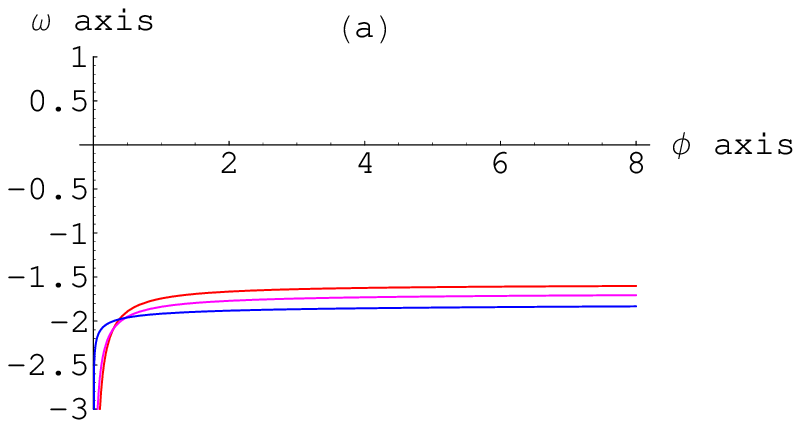,width=.45\linewidth}
\epsfig{file=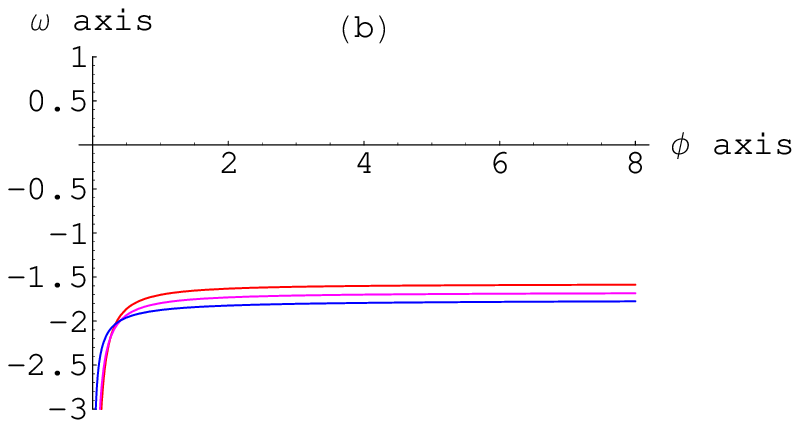,width=.45\linewidth} \caption{Plot of
$\omega(\phi)$ versus $\phi$ for $m=1/2$ and (a) $\gamma=-1/3$ and
(b) $\gamma=-1/2$ with $\alpha>6/5$ as follows: red, $\alpha=1.3$;
blue, $\alpha=2$; pink, $\alpha=1.5$.}
\end{figure}
By taking different choices for these parameters, we see that for
all phases of the universe, the BD parameter $\omega(\phi)$ has
small negative values and lies within the range $\omega\leq -3/2$
as shown in Figures \textbf{8} and \textbf{9} which corresponds to
accelerated expansion of universe. This result is in agreement
with $^{14)}$ for spatially flat model.

\subsection{Model with potential $V(\phi)\neq 0$}

Again, we discuss two cases depending upon the value of BD
parameter $\omega$.

\subsubsection{Case (i)}

First we discuss the case of constant BD parameter, i.e.,
$\omega(\phi)=\omega_{0}$. Further, we consider the power law form
of the scalar field in terms of scale factor $B(t)$
\begin{equation}\label{69}
\phi=\phi_{0}B^{\alpha};~~~\alpha>0.
\end{equation}
Using this value of $\phi$ in the field equations
(\ref{53})-(\ref{55}), it follows that
\begin{equation*}
\frac{\ddot{B}}{B}+A(\frac{\dot{B}^2}{B^2})=-CB^{-\alpha(1+\gamma)(m+2)},
\end{equation*}
where
\begin{equation*}
A=\frac{(\alpha^{2}-3\alpha-2+2m^{2}+\alpha
m-2m+\omega_{0}\alpha^{2})}{(3\alpha+2m)},~~C=\frac{(1+\gamma)\rho_{0}}{\phi_{0}(3\alpha+2m)}.
\end{equation*}
The expression for $\dot{B}(t)$ can be written as
\begin{eqnarray}\nonumber
\dot{B}(t)&=&\sqrt{\frac{2(1+\gamma)\rho_{0}}{\phi_{0}}}\times
B(t)^{\frac{2
-\alpha-(1+\gamma)(m+2)}{2}}\times[(3\alpha+2m)[(1+\gamma)(m+2)\\\label{70}&+&(\alpha-2)]
-2\alpha^{2}+6\alpha +4-4m^{2}-2m\alpha
+4m-2\omega_{0}\alpha^{2}]^{-1/2}
\end{eqnarray}
which yields
\begin{equation}\label{71}
B(t)=A't^{[2/[\alpha+(1+\gamma)(m+2)]},
\end{equation}
where
\begin{eqnarray}\nonumber
A'&=&[\frac{\rho_{0}}{2\phi_{0}}[\alpha+(1+\gamma)(m+2)]^{2}(1+\gamma)\times
[(3\alpha+2m)[(1+\gamma)(m+2)+(\alpha\\\nonumber&-&2)]-2\alpha^{2}+6\alpha
+4-4m^{2}-2m\alpha +4m-2\omega_{0}\alpha^{2}]^{-1}]
^{1/[\alpha+(1+\gamma)(m+2)]}.\\\label{72}
\end{eqnarray}
The value of the scale factor $A(t)$ can be obtained by using
value of $B(t)$ in Eq.(\ref{58}).

The corresponding expression for the scalar field is given by
\begin{equation}\label{72}
\phi(t)=\theta_{0}t^{2\alpha/[\alpha+(1+\gamma)(m+2)]},
\end{equation}
where $\theta_{0}=\phi_{0}A'^{\alpha}$. Equation (\ref{71}) yields
the following constraint on $\alpha$
\begin{equation}\label{73}
3\alpha\leq -(1+3\gamma)(m+2).
\end{equation}
The deceleration parameter $q$ turns out to be
\begin{equation*}
q=-1+\frac{3[\alpha+(1+\gamma)(m+2)]}{2(m+2)}.
\end{equation*}
It can be easily seen that for all positive constant $m$ ($m\neq
1$), $\alpha>1$ and $-1\leq\gamma\leq 1$, the deceleration
parameter remains negative, i.e., $q<0$. Thus the universe is in
the state of accelerated expansion. From the wave equation
(\ref{56}), the potential can be written as
\begin{equation}\label{74}
V(\phi)=\frac{B'}{\phi^{(1+\gamma)(m+2)/\alpha}},
\end{equation}
where $B'$ is given by
\begin{eqnarray}\nonumber
B'&=&\frac{-\alpha\theta_{0}^{2/(\alpha
p)}}{(1+\gamma)(m+2)[\alpha+(1+\gamma)(m+2)]^{2}}[4(1+\gamma)
(m+2)(\omega_{0}\alpha\\\nonumber&-&m-2)
-8(m+2)\alpha\omega_{0}+16m+8m^{2}+24-4m\alpha-8\alpha].
\end{eqnarray}

\subsubsection{Case (ii)}

Let us take the BD parameter as a function of the scalar field
$\phi$, i.e, $\omega(\phi)$. Consider the power law forms for the
scalar field and the scale factor, given by (\ref{60}) and
(\ref{11}). Using the field equations (\ref{53})-(\ref{55}), the
scalar potential takes the form
\begin{eqnarray}\nonumber
V(\phi)&=&\phi_{0}^{(\frac{2}{\beta})}\phi^{\frac{(\beta-2)}{\beta}}[\frac{(m^{2}+5m+6)\alpha^{2}}{2}
-\frac{(m+3)\alpha}{2}+\beta^{2}-\beta+(\frac{(3m+7)}{2})\alpha\beta]
\\\label{75}&-&(1-\gamma)\rho_{0}b_{0}^{-(m+2)(1+\gamma)}\phi^{\frac{-\alpha(m+2)
(1+\gamma)}{\beta}}\phi_{0}^{\frac{\alpha(m+2)(1+\gamma)}{\beta}}.
\end{eqnarray}
The BD parameter turns out to be the same as given by (\ref{67}).
Substituting these values in Eq.(\ref{56}), we obtain the
following consistency relations
\begin{eqnarray}\label{76}
\beta=0,\quad\beta=-2\quad \beta=-\frac{\alpha}{2}(m+3),\quad
\beta=1-\alpha(m+2).
\end{eqnarray}
For the consistency of this solution with the dynamical equation, it
follows that
\begin{equation}\nonumber
\beta=2-\alpha(m+2)(1+\gamma).
\end{equation}

Now we discuss the behavior of self-interacting potential $V(\phi)$
for these values of $\beta$ in different eras of the universe. The
choice $\beta=0$ is not feasible, so we neglect it. For $\beta=-2$,
the self-interacting potential can be written as
\begin{equation*}
V(\phi)=\phi^{2}[(\frac{m^{2}+5m+6}{2})\alpha^{2}-(\frac{m+3}{2})\alpha
+6-(3m+7)\alpha]-(1-\gamma)\phi^{\alpha(m+2)(1+\gamma)/2},
\end{equation*}
where $m\neq1$ is a positive constant and $\alpha>3/(m+2)$. For
$\beta=-\frac{9\alpha}{4},~m=3/2$ and $\alpha>6/7$,  we get
\begin{equation}\nonumber
V(\phi)=\phi^{(1+\frac{8}{9\alpha})}[-3\alpha^{2}]-(1-\gamma)\phi^{14(1+\gamma)/9}.
\end{equation}
For $m=2,~\alpha>3/4$ and $\beta=-\frac{5\alpha}{2}$, the
potential turns out to be
\begin{equation}\nonumber
V(\phi)=-(1-\gamma)\phi^{8(1+\gamma)/5},
\end{equation}
where $-1\leq\gamma\leq 1$.
\begin{figure}
\centering \epsfig{file=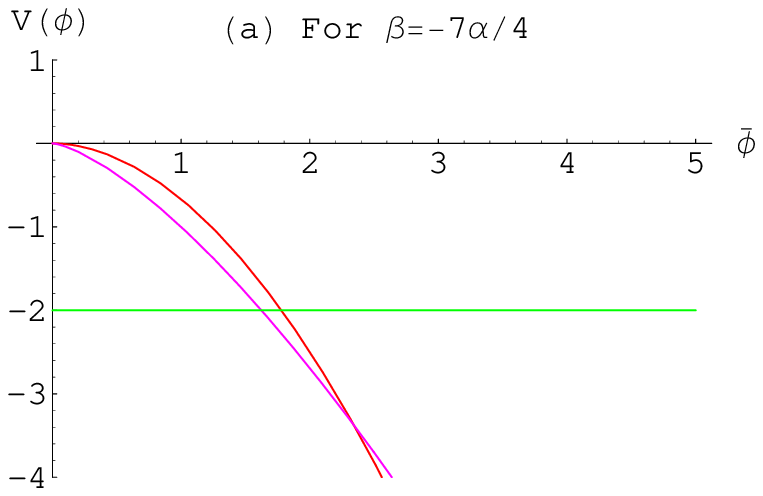,width=.45\linewidth}
\epsfig{file=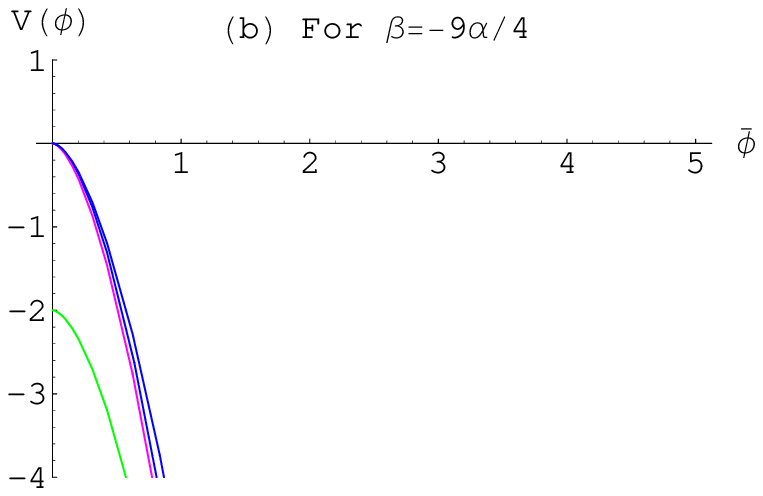,width=.45\linewidth} \caption{Plot (a) shows
the graph of $V(\phi)$ versus scalar field $\phi$ for $m=1/2$,
$\alpha=1.3$ and $\beta=-7\alpha/4$ as follows: red, $\gamma=1/3$;
green, $\gamma=-1$; pink, $\gamma=0$ and Plot (b) shows the graph
of $V(\phi)$ versus scalar field $\phi$ for $m=3/2$, $\alpha=1.3$
and $\beta=-9\alpha/4$ as follows: blue, $\gamma=1/3$; green,
$\gamma=-1$; red, $\gamma=0$; pink, $\gamma=1$.}
\end{figure}
\begin{figure}
\centering \epsfig{file=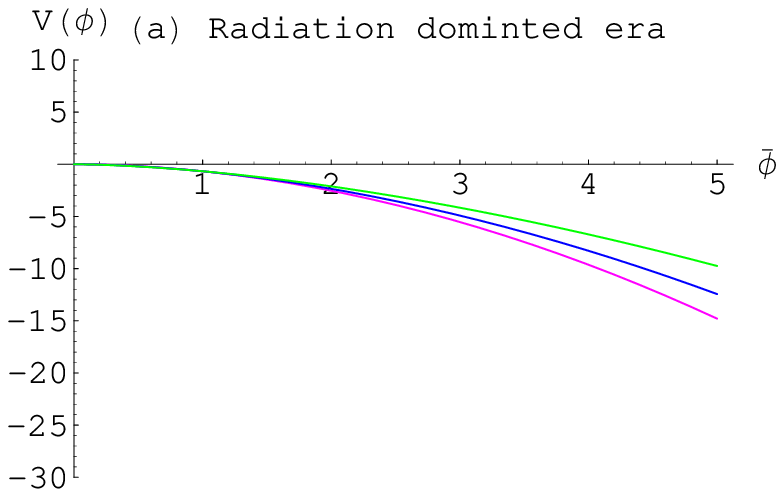,width=.45\linewidth}
\epsfig{file=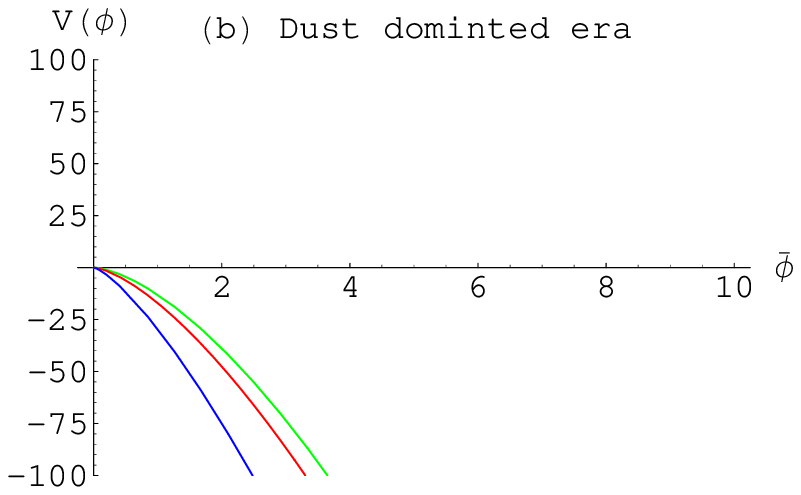,width=.45\linewidth} \caption{Plot (a) shows
the graph of $V(\phi)$ versus scalar field for $m=1/2$,
$\gamma=1/3$ and $\beta=1-5\alpha/2$ as follows: blue,
$\alpha=1.5$; green, $\alpha=2$; pink, $\alpha=1.3$ and Plot (b)
shows the graph of $V(\phi)$ versus scalar field for $m=3/2$,
$\gamma=0$ and $\beta=1-7\alpha/2$ as follows: blue, $\alpha=2$;
green, $\alpha=1.3$; red, $\alpha=1.5$.}
\end{figure}
\begin{figure}
\centering \epsfig{file=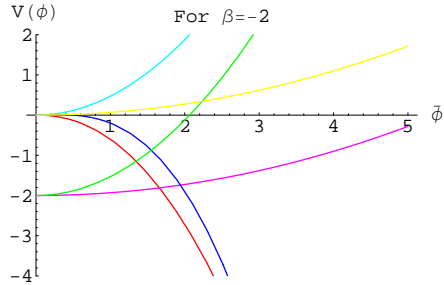,width=.45\linewidth}
\caption{Plot shows the graph of $V(\phi)$ versus scalar field for
$m=1/2$ and $\beta=-2$ as follows: blue, $\alpha=1.5$; red,
$\alpha=1.3$ ($\gamma=1/3$); green, $\alpha=1.5$; pink,
$\alpha=1.3$($\gamma=-1$); sky blue, $\alpha=1.5$; yellow,
$\alpha=1.3$($\gamma=1$).}
\end{figure}
The expression for self-interacting potential for radiation
dominated era with $\beta=2(1-\frac{5\alpha}{3}),~\alpha>6/5$ and
$m=1/2$ is given by
\begin{equation}\nonumber
V(\phi)=(\frac{95\alpha^{2}}{72}-\frac{13\alpha}{4}+\frac{4}{3})\phi^{\frac{-5\alpha/3}{(1-5\alpha/3)}}.
\end{equation}
The self interacting potential for matter dominated era with
$\beta=2-\frac{5\alpha}{2},~\alpha>6/5$ and $m=1/2$ takes the form
\begin{equation}\nonumber
V(\phi)=(1-\frac{3\alpha}{4})\phi^{\frac{-5\alpha/2}{(2-5\alpha/2)}}.
\end{equation}
\begin{figure}
\centering \epsfig{file=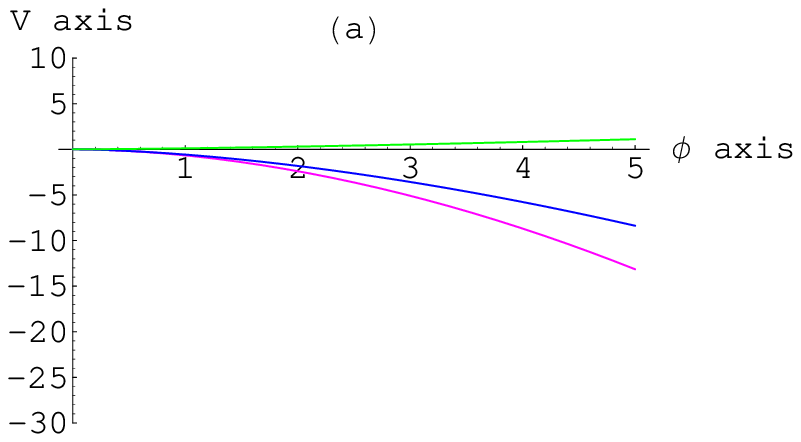,width=.45\linewidth}
\epsfig{file=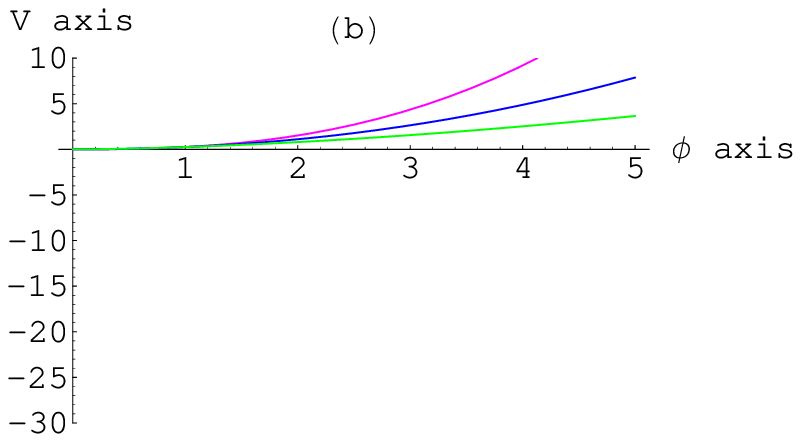,width=.45\linewidth} \caption{Plot (a) shows
the graph of $V(\phi)$ versus scalar field for $m=1/2$,
$\gamma=1/3$ and $\beta=2(1-5\alpha/3)$ as follows: blue,
$\alpha=1.5$; green, $\alpha=2$; pink, $\alpha=1.3$ and Plot (b)
shows the graph of $V(\phi)$ versus scalar field for $m=1/2$,
$\gamma=0$ and $\beta=2-5\alpha/2$ as follows: blue, $\alpha=1.5$;
green, $\alpha=2$; pink, $\alpha=1.3$.}
\end{figure}
For the first three consistency relations for $\beta$ given by
Eq.(\ref{76}), we see that $V(\phi)$ is a decreasing function
starting from zero with the increasing values for $\phi$ except
for the choice $\beta=-2$. In this case, only $\gamma=-1$ and
$\gamma=1$ with $\alpha\geq 1.3$ provide positive potential energy
as for these ranges, they are increasing functions of $\phi$ as
shown in Figures \textbf{10-12}. Figure \textbf{13(a)} shows that
$V(\phi)$ attains negative values starting from zero but with
larger values for $\alpha$, it is an increasing function with
positive values. Figure \textbf{13(b)} shows that $V(\phi)$
attains positive increasing values for $\alpha>6/5$. Therefore, we
conclude that these cases provide positive potential energy as
they are increasing function of $\phi$ for particular values of
$\alpha$.

\section{Variation for the Newton's Gravitational Constant in GBDT}

A well-known fact about BD theory of gravity is that it provides
very small variations for the gravitational constant. However, GBDT
suggests various possibilities for variation of $G$. In GBDT, the
expression for $G$ is found $^{20)}$ to be
\begin{equation}\nonumber
G(t)=\frac{4+2\omega(\phi)}{\phi(3+2\omega(\phi))}.
\end{equation}
The present rate of variation for gravitational constant is given by
\begin{equation}\label{a4}
\left(\frac{\dot{G}}{G}\right)_{0}=-\left(\frac{\dot{\phi}}{\phi}\right)_{0}
-\frac{2(\dot{\omega_{0}})}{(3+2\omega_{0})(4+2\omega_{0})}.
\end{equation}
Here subscript indicates the present values of the corresponding
parameters. Using Eq.(\ref{16}), $\beta=-2\alpha, ~ \alpha>1, ~
\xi_{0}=0.0001$ and the estimated age of the universe $t_{0}=14\pm2$
Gyrs, we obtain the rate of variation of $(\dot{G}/G)_{0}$ to be
$1.5714\times 10^{-19}$ yrs. It lies clearly within the allowed
range of variation of $G$ for cosmic acceleration, that is,
$(\dot{G}/G)_{0}<4\times 10^{-10}$ yrs $^{11,12)}$.

For the Bianchi type $I$ model, by using expression for
$\omega(\phi)$ given by Eq.(\ref{67}) in Eq.(\ref{a4}) along with
values $\beta=-2, ~\alpha>6/5,~ \gamma=0, ~t_{0}=14\pm2$ yrs and
$m=1/2$, we obtain $(\dot{G}/G)_{0}=1.4287\times 10^{-10}$. It also
safely lies within the allowed range of variation of $G$ for cosmic
acceleration. Thus our obtained models satisfy the observational
limit of $G$ for cosmic acceleration.

\section{Summary and Discussion}

This paper investigates the possibility of obtaining cosmic
acceleration by using the role of BD parameter in the presence of
viscous and barotropic fluids. For this purpose, we consider FRW
and BI universe models. The constructed models entirely depend
upon the values of the parameters $\alpha,~\beta,~\gamma$ and $m$.
Firstly, we discuss the FRW model in the presence of viscous
fluid. We see that the total effective pressure contains a
negative factor associated with bulk viscosity which leads to
negative effective pressure. Consequently, the fluid acts as a
dark energy candidate and can explain many aspects of evolution of
the universe. The deceleration parameter constraints the parameter
$\alpha$ for cosmic acceleration i.e., $\alpha>1$.

For the constant coefficient of bulk viscosity, we obtain
$\beta=0$ and $\beta=-2\alpha$. The first case leads to GR while
for the second choice, in all eras of the universe except vacuum
dominated era, the BD parameter $\omega(t)$ is a decreasing
function of time with small negative values. In the vacuum
dominated era, we see that for viscosity greater than $0.11$, it
is possible to achieve cosmic acceleration with positive values of
$\omega(t)$. For the variable bulk viscosity coefficient with
$n=1/2,~\omega(t)$ corresponds to decreasing function as $-t^{2}$
with small negative values for different small values of viscosity
coefficient $\xi_{0}$ and $-1\leq\gamma\leq 1$. This gives rise to
accelerated expansion of the universe for $\alpha>1$. For the
radiative fluid, we have found that the coefficient of viscosity
appears in exponential function. Here $\omega(t)$ is a decreasing
function with negative values both for the accelerated and
decelerated phases of the universe ($-3/2\leq\omega\leq0$).

Secondly, we have taken the BI universe model in the presence of
perfect fluid with barotropic EoS. Here we have taken two cases
when $V(\phi)=0$ and  $V(\phi)\neq 0$. In the first case, when
$\omega(\phi)=\omega_{0}$, by taking different possible choices
for the parameters, we see that the BD parameter takes small
negative as well as positive values for $-1\leq\gamma<0$ and
certain ranges of $\alpha$. Thus the cosmic acceleration can be
achieved for positive larger values of $\omega$ with different
values of $\alpha$. Also, for the present universe with
$\omega_{0}=-5/3$ (taken from the negative observed range for
cosmic acceleration $-2\leq\omega_{0}\leq -3/2$), we have found
$\phi(t)\sim t^{-3}$. In this case, the acceleration rate of the
universe is higher than Bertolami et al. $(\phi(t)\sim t^{-2})$
$^{11)}$ and Benerjee et al. $(\phi(t)\sim t^{-5/2})$ $^{12)}$.
When $\omega=\omega(\phi)$, taking different values of the
parameters, we see that for all phases of the universe, the BD
parameter $\omega(\phi)$ have small negative values and lies
within the range $\omega\leq -3/2$ which corresponds to cosmic
acceleration and in agreement with already found results $^{14)}$.

For $V(\phi)\neq 0$ and $\omega(\phi)=\omega_{0}$, we have evaluated
the values of scale factors $A$, $B$, scalar field and $V(\phi)$. It
is found that in all phases of the universe, these values of scale
factors $A(t)$ and $B(t)$ lead to $q<0$ for all positive constant
$m$ with $m\neq 1$ and $\alpha>1$ which corresponds to cosmic
acceleration. Finally, for $V(\phi)\neq 0$ and
$\omega=\omega(\phi)$, we see that $V(\phi)$ is a decreasing
function starting from zero with the increasing $\phi$ except for
the choice $\beta=-2$ with particular values of $\gamma$ and
$\alpha\geq 1.3$. These values provide positive potential energy as
they are increasing function of $\phi$. However, for the constraint
$\beta=2-\alpha(m+2)(1+\gamma)$, it is possible to have positive
potential energy for larger values of $\alpha$ in matter dominated
era while for smaller values of $\alpha$ in radiation dominated era.

It is worthwhile to mention here that all models discussed here
satisfy the observational constraints for the variation of Newton's
gravitational constant available in literature $^{11,12)}$ which
provides a support to our obtained results. Although in each case,
we have explained the phenomena of cosmic acceleration for different
ranges of the corresponding parameters. However, these ranges of the
BD parameter, except for few cases, are incompatible with solar
system constraints which require $\omega\geq 40,000$. This is the
generic problem noted in the context of scalar tensor theories. It
would be of great interest to see whether this problem can be
resolved using other Bianchi models.

In order to check the viability of dark energy models based on
modified theories of gravity, the evolution of cosmological
perturbations and the background expansion history of the universe
may be studied. This can be done through Chameleon and Vainshtein
mechanisms which suppress the propagation of fifth force and provide
consistency with local gravity experiments $^{50,51)}$. One may
adopt these procedures to check the viability of above discussed
models.\\\\

1) S. Perlmutter, S. Gabi, G. Goldhaber, A. Goobar, D. E. Groom, I.
M. Hook, A. G. Kim, M. Y. Kim, G. C. Lee, R. Pain, C. R.
Pennypacker, I. A. Small, R. S. Ellis, R. G. McMahon, B. J. Boyle,
P. S. Bunclark, D. Carter, M. J. Irwin, K. Glazebrook, H. J. M.
Newberg, A. V. Filippenko, T. Matheson, M. Dopita and W. C. Couch:
Astrophys. J. \textbf{483}(1997)565; S. Perlmutter, G. Aldering, M.
D. Valle, S. Deustua, R. S. Ellis, S. Fabbro, A. Fruchter, G.
Goldhaber, A. Goobar, D. E. Groom, I. M. Hook, A. G. Kim, M. Y. Kim,
R. A. Knop, C. Lidman, R. G. McMahon, P. Nugent, R. Pain, N.
Panagia, C. R. Pennypacker, P. Ruiz-Lapuente, B. Schaefer and N.
Walton: Nature \textbf{391}(1998)51; S. Perlmutter, G. Aldering, G.
Goldhaber, R. A. Knop, P. Nugent, P. G. Castro, S. Deustua, S.
Fabbro, A. Goobar, D. E. Groom, I. M. Hook, A. G. Kim, M. Y. Kim, J.
C. Lee, N. J. Nunes, R. Pain, C. R. Pennypacker, R. Quimbey, C.
Lidman, R. S. Ellis, M. Irwin, R. G. Mcmahon, P. Ruiz-lapuente, N.
Walton, B. Schaefer, B. J. Boyle, A. V. Filippenko, T. Matheson, A.
S. Fruchter, N. Panagia, H. J. M. Newberg and
W. J. Couch: Astrophys. J. \textbf{517}(1999)565.\\
2) A. G. Riess, A. V. Filippenko, P. Challis, A. ClocChiatti, A.
Diercks, P. M. Garnavich, R. L. Gilliland, C. J. Hogan, S. Jha, R.
P. Kirshner, B. Leibundgut, M. M. Phillips, D. Reiss, B. P. Schmidt,
R. A. Schommer, R. C. Smith, J. Spyromilio, C. Stubbs, N. B.
Suntzeff and J. Tonry: Astron. J. \textbf{116}(1998)1009.\\
3) C. L. Bennett, M. Halpern, G. Hinshaw, N. Jarosik, A. Kogut, M.
Limon, S. S. Meyer, L. Page, D. N. Spergel, G. S. Tucker, E.
Wollack, E. L. Wright, C. Barnes, M. R. Greason, R. S. Hill, E.
Komatsu, M. R. Nolta, N. Odegard, H. V. Peiris, L. Verde and
J. L. Weiland: Astrophys. J. Suppl. \textbf{148}(2003)1.\\
4) M. Tegmark, M. A. Strauss, M. R. Blanton, K. Abazajian, S.
Dodelson, H. Sandvik, X. Wang, D. H. Weinberg, I. Zehavi, N. A.
Bahcall, F. Hoyle, D. Schlegel, R. Scoccimarro, M. S. Vogeley, A.
Berlind, T. Budavari, A. Connolly, D. J. Eisenstein, D. Finkbeiner,
J. A. Frieman, J. E. Gunn, L. Hui, B. Jain, D. Johnston, S. Kent, H.
Lin, R. Nakajima, R. C. Nichol, J. P. Ostriker, A. Pope, R.
Scranton, U. Seljak, R. K. Sheth, A. Stebbins, A. S. Szalay, I.
Szapudi, Y. Xu, J. Annis, J. Brinkmann, S. Burles, F. J. Castander,
I. Csabai, J. Loveday, M. Doi, M. Fukugita, B. Gillespie, G.
Hennessy, D. W. Hogg, Z. E. Ivezic´, G. R. Knapp, D. Q. Lamb, B. C.
Lee, R. H. Lupton, T. A. McKay, P. Kunszt, J. A. Munn, L. Connell,
J. Peoples, J. R. Pier, M. Richmond, C. Rockosi, D. P. Schneider, C.
Stoughton, D. L. Tucker, D. E. V. Berk, B. Yanny and D. G. York:
Phys. Rev. \textbf{D69}(2004)03501.\\
5) S. W. Allen, R. W. Schmidt, H. Ebeling, A. C. Fabian and L. V.
Speybroeck: Mon. Not. Roy. Astron. Soc.
\textbf{353}(2004)457.\\
6) E. Hawkins, S. Maddox, S. Cole, O. Lahav, D. S. Madgwick, P.
Norberg, J. A. Peacock, I. K. Baldry, C. M. Baugh, J.
Bland-Hawthorn, T. Bridges, R. Cannon, M. Colless, C. Collins, W.
Couch, G. Dalton, R. D. Propris, S. P. Driver, S.P., G. Efstathiou,
R. S. Ellis, C.S. Frenk, K. Glazebrook, C. Jackson, B. Jones, I.
Lewis, S. Lumsden, W. Percival, B. A. Peterson, W. Sutherland and K.
Taylor: Mon. Not. Roy. Astr. Soc.
\textbf{346}(2003)78.\\
7) B. Jain and A. Taylor: Phys. Rev. Lett.
\textbf{91}(2003)141302.\\
8) R. R. Caldwell, R. Dave and P. J. Steinhardt: Phys. Rev. Lett.
\textbf{80}(1998)1582.\\
9) A. S. Al-Rawaf and M. O. Taha: Gen. Relativ. Gravit.
\textbf{28}(1996)935.\\
10) A. H. Guth: Phys. Rev. \textbf{D23}(1981)347.\\
11) O. Bertolami and P. J. Martins: Phys. Rev.
\textbf{D61}(2000)064007.\\
12) N. Banerjee and D. Pavon: Phys. Rev.
\textbf{D63}(2001)043504.\\
13) B. K. Sahoo and L. P. Singh: Mod. Phys. Lett.
\textbf{A18}(2003)2725.\\
14) W. Chakraborty and U. Debnath: Int. J. Theor. Phys.
\textbf{48}(2009)232.\\
15) C. H. Brans and R. H. Dicke: Phys. Rev. \textbf{124}(1961)925.\\
16) S. Weinberg: \emph{Gravitation and Cosmology} (Wiley, 1972).\\
17) P. A. M. Dirac: Proc. R. Soc. Lond. \textbf{A165}(1938)199.\\
18) B. Bertotti, L. Iess and P. Tortora: Nature
\textbf{425}(2003)374.\\
19) A. D. Felice, G. Mangano, P. D. Serpico and M. Trodden: Phys. Rev. \textbf{D74}(2006)103005.\\
20) K. Nordtvedt Jr.: Astrophys. J. \textbf{161}(1970)1059. \\
21) R. V. Wagoner: Phys. Rev. \textbf{D1}(1970)3209.\\
22) T. Singh and L. N. Rai: Gen. Relativ. Gravit.
\textbf{B15}(1983)875.\\
23) M. S. Bermann:  Nuovo Cimento \textbf{B74}(1983)192.\\
24) S. Sen and A. A. Sen: Phys. Rev. \textbf{D63}(2001)124006.\\
25) B. K. Sahoo and L. P. Singh: Mod. Phys. Lett.
\textbf{A17}(2002)2409.\\
26) S. Sen and T. R. Seshadri: Int. J. Mod. Phys.
\textbf{D12}(2003)445.\\
27) D. R. K. Reddy and M. V. S. Rao: Astrophys. Space. Sci.
\textbf{305}(2006)183.\\
28) J. P. Singh and P. S. Baghel: Elect. J. Theor. Phys.
\textbf{6}(2009)85.\\
29) M. K. Verma, M. Zeyauddin and S. Ram: Rom. J. Phys.
\textbf{56}(2011)616.\\
30) M. Sharif and S. Waheed: Eur. Phys. J. \textbf{72C}(2012)1876;
A. K. Yadav and B. Saha: Astrophys. Space. Sci \textbf{337}(2012)759.\\
31) C. P. Singh: Brazil. J. Phys. \textbf{39}(2009)619.\\
32) S. Fay: Gen. Relativ. Gravit. \textbf{32}(2000)203.\\
33) S. K. Rama and S. Gosh: Phys. Lett. \textbf{B383}(1996)32;
S. K. Rama: Phys. Lett. \textbf{B373}(1996)282.\\
34) C. Romero and A. Barros: Phys. Lett. \textbf{A173}(1993)243.\\
35) N. Banerjee and S. Sen: Phys. Rev. \textbf{D56}(1997)1334.\\
36) A. Bhadra and K. K. Nandi: Phys. Rev.
\textbf{D64}(2001)087501.\\
37) C. H. Brans: Phys. Rev. \textbf{125}(1962)2194.\\
38) T. P. Sotiriou and V. Faraoni: Rev. Mod. Phys.
\textbf{82}(2010)451.\\
39) C. Eckart: Phys. Rev. \textbf{58}(1940)919.\\
40) G. L. Murphy: Phys. Rev. \textbf{D8}(1973)4231.\\
41) S. Weinberg: Astrophys. J. \textbf{168}(1971)175.\\
42) U. A. Belinskii and I. M. Khalatnikov: Sov. Phys. JETP
\textbf{42}(1976)205.\\
43) G. P. Singh, S. G. Ghosh and A. Beesham: Aust. J. Phys.
\textbf{50}(1997)903.\\
44) M. Sharif and M. Zubair: Astrophys. Space Sci.
\textbf{330}(2010)399.\\
45) C. B. Collins, E. N. Glass and D. A. Wilkinson: Gen. Relativ.
Gravit. \textbf{12}(1980)805.\\
46) K. S. Throne: Astrophys. J. \textbf{148}(1967)51.\\
47) J. Kristian and R. K. Sachs: Astrophys. J.
\textbf{143}(1966)379.\\
48) C. B. Colins: Phys. Lett. \textbf{A60}(1977)397.\\
49) S. R. Roy and S. K. Banerjee: Class. Quantum Grav.
\textbf{11}(1995)1943.\\
50)  G. Radouane, M. Bruno, F. M. David, P. David, T. Shinji and
A. W. Hans: Phys. Rev. \textbf{D82}(2010)124006.\\
51) T. Shinji: Lect. Notes Phys. \textbf{800}(2010)99.
\end{document}